\documentclass[12pt]{article}
\usepackage{newtxtext,newtxmath}
\usepackage{graphicx}
\usepackage[letterpaper,margin=1in]{geometry}
\renewenvironment{abstract}
	{\quotation}
	{\endquotation}

\date{}

\makeatletter
\renewcommand{\fnum@figure}{\textbf{Figure \thefigure}}
\renewcommand{\fnum@table}{\textbf{Table \thetable}}
\makeatother

\usepackage{scicite}

\usepackage{url}

\def\scititle{
	A Constructed Closure of the Bering Strait can Prevent an AMOC Tipping
}
\title{\bfseries \boldmath \scititle}

\author{
	Jelle~Soons$^{1\ast}$ and Henk~A.~Dijkstra$^{1}$\and
	\small$^{1}$Institute for Marine and Atmospheric Research Utrecht, Utrecht University,\and 
    \small Princetonplein 5, Utrecht 3584 CC, The Netherlands.\and
	\small$^\ast$Corresponding author. Email: j.soons@uu.nl\and
}

\begin{document} 

\maketitle



\begin{abstract} \bfseries \boldmath
The Atlantic Meridional Overturning Circulation (AMOC) is a major tipping element in the present-day climate, and could potentially 
collapse under sufficient freshwater or CO$_2$-forcing. While the effect of the Bering Strait on AMOC stability has been well studied, it 
is unknown whether a constructed closure of this Strait can prevent an AMOC collapse under climate change.  Here, we show in an 
Earth system Model of Intermediate Complexity that an artificial closure of the Strait can extend the safe carbon budget of 
the AMOC, provided that the AMOC is strong enough at the closure time. Specifically, for this model, an equilibrium AMOC with a reduction below  $(6.1\pm0.5)\%$ from pre-industrial has an additional budget up to $500$~PgC 
given a sufficiently early closure, while for a weaker AMOC a closure reduces this budget. This indicates that constructing this closure can be a feasible climate intervention strategy to prevent an AMOC collapse.
\end{abstract}

\subsection*{Main}
The Atlantic Meridional Overturning Circulation (AMOC) is of paramount importance in regulating Earth's climate. It transports warm surface waters from the tropics northward, which is a major reason for the relatively mild climate in Europe despite its hight latitude~\cite{srokosz2015observing, buckley2016observations}. A key driving factor in this overturning circulation is the water mass transformation in the Nordic Seas, where the relatively warm and salty surface waters are cooled by the atmosphere, and become the cold and salty North Atlantic Deep Water (NADW). This dense water mass sinks, and returns southward~\cite{frajka2019atlantic}. 
There is a growing concern that under global warming the AMOC could weaken or even shut down~\cite{manabe1993century, weijer2020cmip6, bellomo2021future, caesar2021current, boers2021observation}, with some studies even warning for an oncoming collapse this century~\cite{ditlevsen2023warning, van2025phys} while others have ruled out this possibility~\cite{baker2025continued,bonan2025observational}. A possible collapse would have a major impact on the global climate, particularly Europe's~\cite{hirota2011global, van2024physics, van2025european}, and could be practically irreversible as multiple equilibrium states have been found consistently throughout the model hierarchy~\cite{stommel1961thermohaline, rahmstorf2005thermohaline, hawkins2011bistability, van2023asymmetry, dijkstra2024role}. 

A non-negligible effect on the AMOC's stability is the Bering Strait Throughflow (BST).  Almost $1$~Sv of North Pacific surface water flows northward through the Bering Strait, where it eventually ends up in the Labrador and Greenland Sea joining the lower limb of the AMOC 
~\cite{yang2024bering}. This water is relatively fresh ($\sim 32.5$ g/kg) as it originates from Antarctic Intermediate Water, and the net northward freshwater transport is roughly $80$~mSv~\cite{haine2015arctic}. Consequently, it inhibits the northern deep water formation and in turn weakens the AMOC. Paleoclimate model simulations show that a closure of the Bering Strait (CBS) leads to a stronger AMOC with increased meridional heat transport~\cite{goosse1997sensitivity, hu2015effects, otto2017amplified, weiffenbach2025impact}. Not only because it prevents fresh North Pacific waters from entering, but also because it reduces upper ocean water exchange between the Arctic and North Atlantic, which also reduces the input of fresh water into the North Atlantic~\cite{goosse1997sensitivity, otto2017amplified}. Under present-day climate conditions a closure would increase the AMOC strength by $2.5\pm0.5$~Sv~\cite{hu2015effects}. 

Although the BST plays only a minor role in an AMOC collapse it can significantly affect AMOC stability under freshwater flux forcing, also 
called hosing ~\cite{hu2012role, vanderborght2024feedback}. Climate model simulations in ~\cite{hu2012role} indicate that for low hosing 
values the AMOC strength is higher under CBS, but as the hosing increases the AMOC declines more rapidly than under an open Bering Strait (OBS) resulting in the AMOC's critical hosing value being lower for a CBS. As the AMOC decreases,  the sea-level in the Arctic increases, which causes the BST to decrease or even reverse~\cite{cessi2020control, van2024physics}. This produces an export of the added freshwater to the Pacific. Hence, the BST functions as a stabilizing mechanism for an active AMOC, and as a destabilizing mechanism for a collapsed AMOC state~\cite{de2004bering, hu2005bering, hu2012role, hu2015effects}. Interestingly, this stabilizing effect of the BST on a freshwater forced 
AMOC is not as pronounced under CO$_2$ forcing due to additional changes in the hydrological cycle~\cite{hu2023dichotomy}. Under 
CO$_2$ forcing,  the AMOC weakens due to increased heating instead of freshening of the North Atlantic's surface.

\begin{figure}
    \centering
    \includegraphics[width = \textwidth]{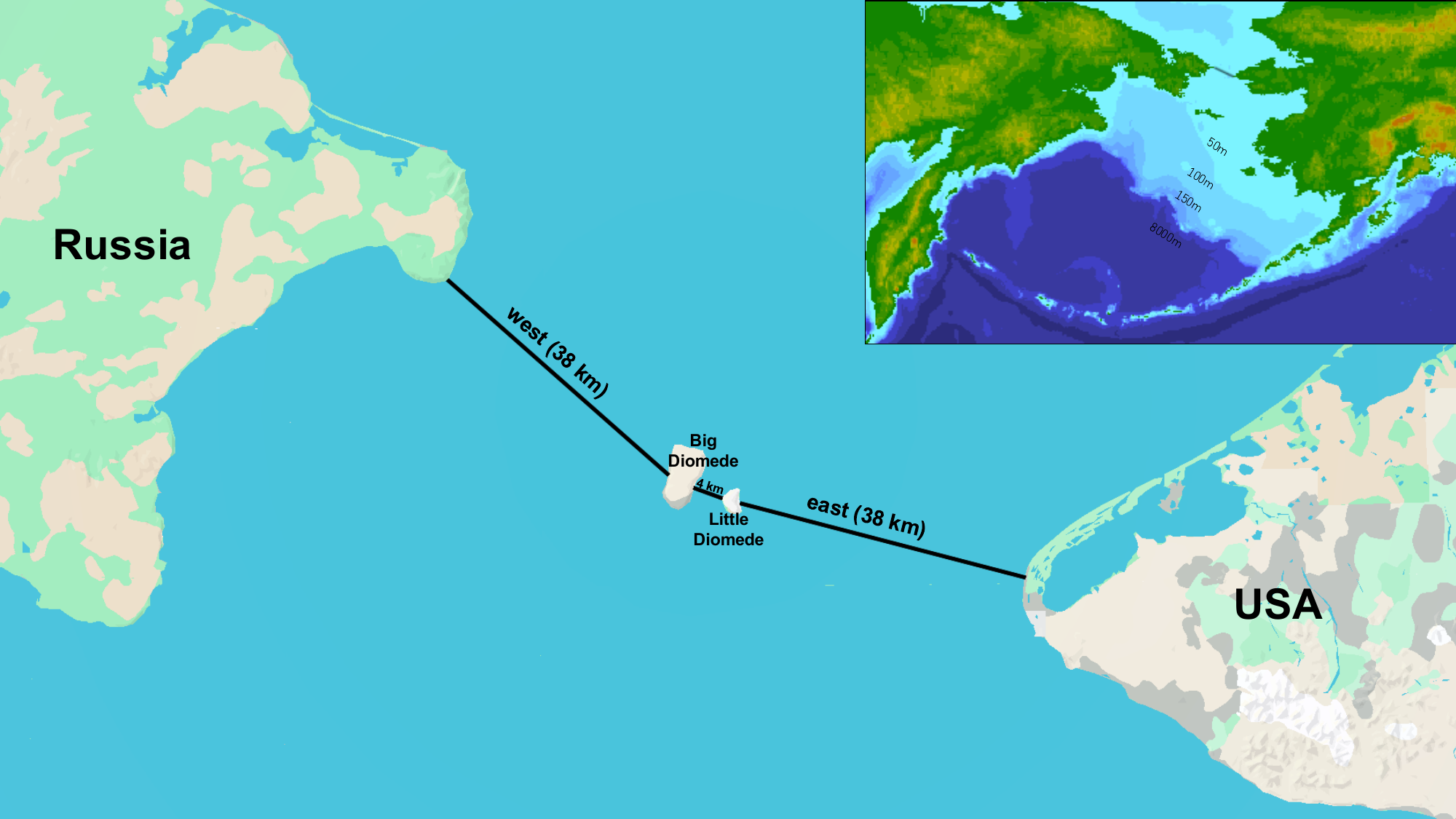}
    \caption{\textbf{The Bering Strait Dam.} The proposed Bering Strait Dam (BSD, thick black lines) consisting of three separate dams: a western section connecting mainland Russia with Big Diomede Island ($\sim 38$ km), a middle section connecting Big Diomede Island to Little Diomede Island ($\sim 4$ km), and an eastern section connecting Little Diomede Island to Alaska, USA ($\sim 38$ km). The inset map indicates the bathymetry of the Bering Sea with the BSD added [adjusted from~\cite{nasa_beringsea}]. Both maps are oriented with north at the top.}
    \label{fig:BSDmap}
\end{figure}

Observation based early warning signals are indicating that the AMOC looses resilience ~\cite{boers2021observation} and  
that an onset of an AMOC collapse may occur over the next decades ~\cite{ditlevsen2023warning, van2025phys}. The consequences  are 
substantial ~\cite{van2025european, rahmstorf2024atlantic} and strongly motivate to consider feasible intervention strategies 
to prevent such a  AMOC collapse. Both carbon dioxide removal (CDR) techniques and Solar Radiation 
Management (SRM) can be effective tools to limit global warming and prevent an AMOC collapse, but their deployment 
come with significant technical, economical and governance considerations~\cite{geden2023state, powis2023quantifying, 
chiquier2025integrated, aaheim2015economic, irvine2016overview}. We here propose as an intervention the construction 
of a Bering Strait Dam (BSD). The BSD would disconnect the Pacific Ocean from the Arctic Ocean with three separate 
dams (Figure~\ref{fig:BSDmap}). It consists of a western section connecting mainland Russia to Big Diomede Island, a 
middle section connecting the Diomede Islands, and an eastern section connecting Little Diomede Island to Alaska, USA. 
Combined these sections have a length of roughly $80$~km and encounter an average depth of $50$~m with a maximum 
depth of $59$~m~\cite{borisov1969can, cathcart2011bering}. Given these dimensions the construction of the BSD is 
considered to be technically feasible~\cite{groeskamp2020need}.

\subsection*{Results}

We study the AMOC's response to CO$_2$-forcing and the influence of a Bering Strait closure on this response using 
the Earth system model CLIMBER-X~\cite{willeit2022earth, willeit2023earth} (see Methods).  All components of the 
climate model have a horizontal resolution of $5^\circ\times 5^\circ$. Earlier work has shown that there are four distinct 
AMOC equilibrium states in CLIMBER-X: a strong AMOC state, a modern AMOC state,  a weak AMOC state, and a 
collapsed AMOC~\cite{willeit2024generalized}. We will refer to the first two both as an ON state, and  the latter two 
as an OFF state. Note that in this model the produced AMOC ON state is slightly too shallow and lacks a deep 
southern overturning cell, while the OFF state lacks a strong reversed circulation near the surface~\cite{willeit2022earth, 
van2023asymmetry, van2024physics}, see also Figure~\ref{fig:supAMOC}.   At this relatively low resolution we can 
simulate  almost $10,000$ model years per day, allowing us to do many simulations in order to find the safe carbon 
budget  under different background conditions. To understand the stability of the AMOC under CO$_2$ forcing, we 
need to consider first its sensitivity to freshwater forcing. 

\subsubsection*{Freshwater forcing}
To determine the stability of the AMOC in CLIMBER-X under OBS and CBS under freshwater flux forcing, we  
perform two quasi-equilibrium hysteresis experiments.  Here a slowly varying freshwater flux forcing with 
strength $F_H$ is applied between latitudes $20^\circ$N and $50^\circ$N in the Atlantic. This region is a 
common choice in AMOC hysteresis experiments~\cite{rahmstorf2005thermohaline, hu2012role, van2023asymmetry}, 
and yields a straightforward  hysteresis curve~\cite{boot2025physics, willeit2024generalized}. The freshwater flux is 
compensated over the  rest of the  ocean domain. The initial state is an equilibrium AMOC state under  pre-industrial 
conditions, i.e. no hosing ($F_H = 0$  Sv)  and the CO$_2$ concentration is fixed at $280$~ppm.  The freshwater 
flux $F_H$ is increased linearly in time at a rate of $0.025$~Sv/kyr until a total flux of $0.35$ Sv is achieved, after 
which the flux is reduced with the same rate till the hosing flux is $-0.25$~Sv. Then finally, the hosing flux is increased 
to its original zero level. This rate was based on those used for quasi-equilibrium simulations in previous studies on 
AMOC hysteresis curves using freshwater hosing in CLIMBER-X~\cite{willeit2024generalized, boot2025physics}.

Throughout this study we use the notation $\Delta\,Q$ to indicate the difference in quantity $Q$ between the CBS and OBS conditions, i.e. $\Delta\,Q = Q_{\text{CBS}} - Q_{\text{OBS}}$.   The AMOC strength at $26^\circ$N under CBS is  stronger up to a hosing value $F_H = 0.161$~Sv with an overturning strength of $19.9$~Sv versus $19.6$~Sv (CBS versus OBS) for $F_H = 0$ Sv  (Figure~\ref{fig:hysteresis}\textbf{A, B}). This difference is an order of magnitude smaller compared to previous studies using more detailed climate models~\cite{goosse1997sensitivity, hu2015effects}. With the closure, the critical hosing value or an AMOC  collapse is lowered to $0.195$~Sv as opposed to $0.220$~Sv under 
OBS. Moreover, the AMOC OFF state is much more stable under CBS, as its recovery occurs for a hosing value that is roughly $0.13$~Sv 
lower. This also produces the larger overshoot during this recovery under CBS as a larger amount of salinity is suddenly transported 
northward. 

The difference in surface density and salinity is rather subtle for the ON-states, with initially a 
higher density and salinity under CBS for lower hosing values (Figure~\ref{fig:hysteresis}\textbf{C, D}). Here the surface layer of the North Atlantic is the top $200$~m layer between $50^\circ$ and $75^\circ$N in the Atlantic basin, see also Materials and Methods, and figure~\ref{fig:supregion}.  As the hosing flux increases the difference switches sign, with both the density and salinity under CBS dipping below those under OBS, starting at $F_H = 0.164$~Sv. For the OFF-states,  a closure results in a much lower surface density and salinity. This explains the increased stability of the AMOC OFF state under closure. All in all, a closure does increase the surface density of the North Atlantic --and correspondingly the AMOC strength-- for low freshwater forcing, but has the reversed effect for higher freshwater forcing and so results in a lower critical hosing value. As the difference in density is mainly explained by the difference in salinity (Figure~\ref{fig:hysteresis}\textbf{D}), the difference in AMOC behavior is caused here by a difference in freshwater transports into the North Atlantic region.

\begin{figure}
    \centering
    \includegraphics[scale = 0.4]{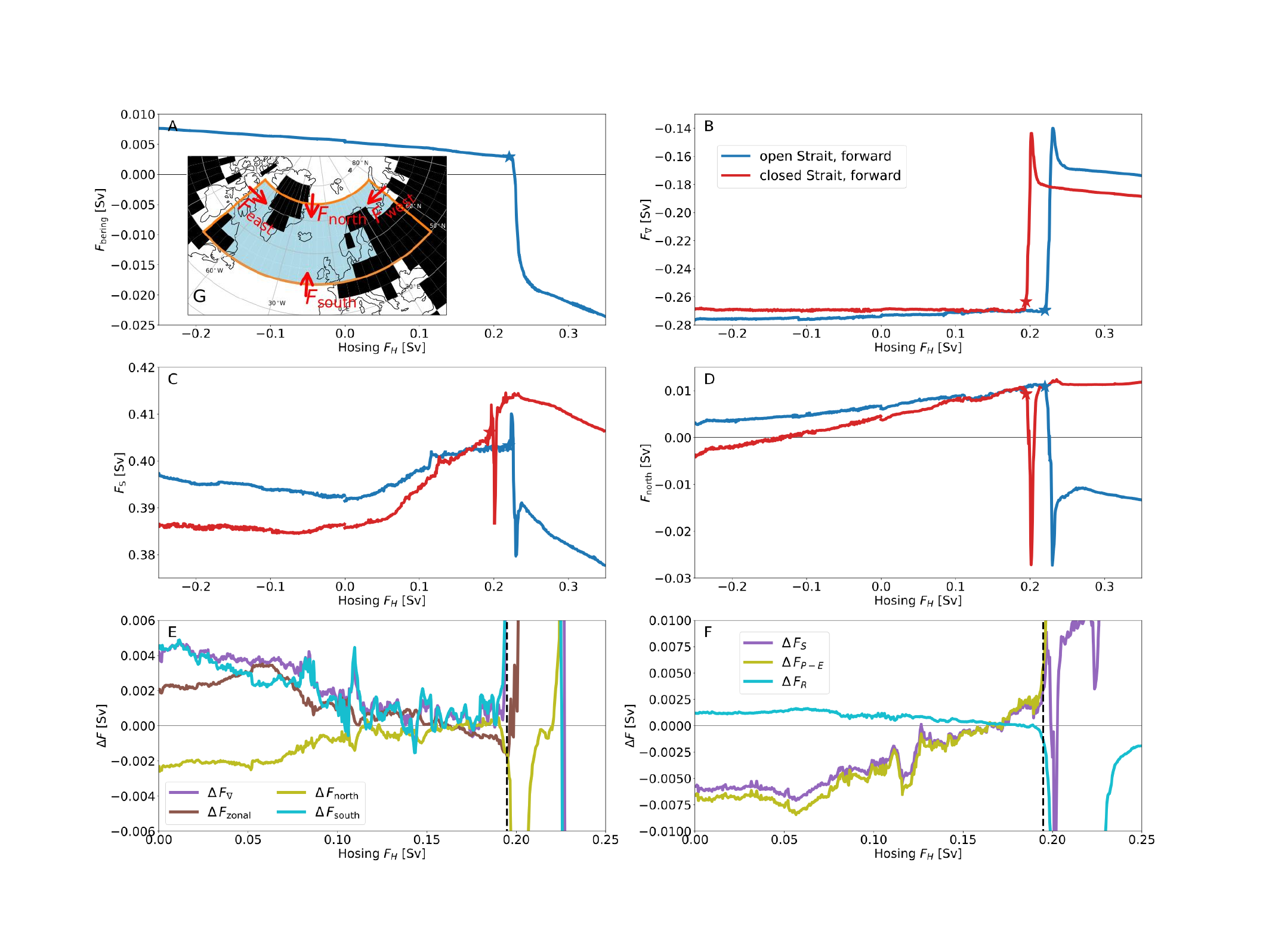}
    \caption{\textbf{The freshwater transports.} Freshwater transports for the quasi-equilibrium simulations for an open Strait (blue), and a closed Strait (red), consisting of simulations where the hosing flux $F_H$ increases (\textbf{A}-\textbf{D}). The asterisks mark the estimated tipping points of the AMOC collapses (\textbf{A}-\textbf{D}), while the vertical line (dashed, black) indicates the critical hosing value for the AMOC tipping under CBS (\textbf{E}\&\textbf{F}). (\textbf{A-D}) The freshwater transports through, respectively, the Bering Strait ($F_{\text{bering}}$), the lateral boundaries of the North Atlantic region ($F_{\nabla}$), the surface of the North Atlantic region ($F_S$), and the northern boundary of the North Atlantic region ($F_{\text{north}}$). (\textbf{E}) The difference between the ON-states under CBS and OBS in the freshwater transports through the the North Atlantic's boundaries ($\Delta\,F_{\nabla}$, purple), its zonal boundaries ($\Delta\,F_{\text{zonal}}$, brown), its northern boundary ($\Delta\,F_{\text{north}}$, citrus), and its southern boundary ($\Delta\,F_{\text{south}}$, cyan). (\textbf{F}) The difference between the ON-states under CBS and OBS in the freshwater transports through the North Atlantic's surface ($\Delta\,F_S$, purple), consisting mainly of the differences in precipitation-minus-evaporation ($\Delta\,F_{P-E}$, citrus) and in runoff ($\Delta\,F_R$, cyan). Inset (\textbf{G}) indicates the lateral freshwater transports and their direction (red arrows) into the North Atlantic region (light-blue, enclosed in orange). The black cells indicate grid cells with a zero ocean fraction on top of the current coastlines (black, solid).} 
    \label{fig:fwtransport}
\end{figure}

As it is important to understand the AMOC stability under CO$_2$ forcing below, we consider these freshwater transports (see Methods) 
in more detail for the forward runs (i.e. the simulations with slowly increasing hosing) over the North Atlantic region shown in Figure~\ref{fig:supregion}. Those for the backward runs are discussed in the supplementary materials. 
Note that all freshwater transports are positive when directed into the region, see also Figure~\ref{fig:fwtransport}\textbf{G}. For OBS the freshwater transport out off the North Pacific into the 
Arctic ($F_{\text{bering}}$ in Figure~\ref{fig:fwtransport}\textbf{A}) reduces as the AMOC weakens from $5.4$~mSv to $2.9$~mSv at 
the tipping point, and reverses for  a collapsed AMOC to a net southward freshwater transport as high as $18.4$~mSv. This  is 
qualitatively consistent with previous studies~\cite{hu2005bering, hu2012role, hu2015effects, otto2017amplified} and theory~\cite{de2004bering,cessi2020control},  although it is an order of magnitude smaller than current observations
~\cite{haine2015arctic}.    

The effect of the weakening AMOC on the freshwater exchange to the North Atlantic from the rest of the ocean basin ($F_{\nabla}$), and from the atmosphere, lithosphere and cryosphere ($F_S$), and from the Arctic Ocean ($F_{\text{north}}$) can be seen in Figure~\ref{fig:fwtransport}\textbf{B}-\textbf{D}, respectively.  Under an active AMOC there is a net freshwater export out off the North Atlantic through its lateral boundaries, ($F_{\nabla} < 0$, Figure~\ref{fig:fwtransport}\textbf{B}) which is slightly larger for OBS. Under a collapsed AMOC this net freshwater export has decreased, and is significantly smaller under OBS. On the other hand, with a collapsed AMOC the freshwater transport from the Arctic to the North Atlantic is southward under CBS (Figure~\ref{fig:fwtransport}\textbf{D}), while northward under OBS, explaining the much fresher North Atlantic under CBS, and hence the more stable OFF-state. The difference $\Delta\,F_{\text{north}}$ (green curve in Figure~\ref{fig:fwtransport}\textbf{E}) for an AMOC ON state is much more subtle, with a slightly larger southward freshwater transport under OBS, before the difference is negligible for $F_H\gtrapprox 0.10$~Sv up to the  tipping point. Lastly, the surface flux $F_S$ (Figure~\ref{fig:fwtransport}\textbf{C}) onto the North Atlantic for an active AMOC is larger under OBS for lower hosing values, but the roles reverse for larger hosing $F_H>0.16$~Sv. Together with the freshwater transport from the Arctic this reversal explains the  more saline North Atlantic --and hence stronger AMOC-- for low hosing values under CBS, and the flip for higher hosing values. The surface forcing during a collapsed AMOC is lower under CBS than under OBS, and does not explain the fresher North Atlantic. 

Figures~\ref{fig:fwtransport}\textbf{E}\&\textbf{F} also display the other differences between the various freshwater transports into the North Atlantic under CBS and OBS conditions. As discussed, for low hosing values there is a larger freshwater import under CBS ($\Delta\,F_{\nabla}>0$), which is caused by a larger import through the southern and zonal boundaries. At the same time there is a larger export through the northern boundary under CBS ($\Delta\,F_{\text{north}}>0$). All these differences slowly vanish for increased hosing flux. For the surface flux we find  a larger freshwater input to the ocean's surface under OBS conditions ($\Delta\,F_{S}<0$), which reverses for increased hosing. This is mainly due to the difference in the precipitation-minus-evaporation above the North Atlantic ($\Delta\,F_{P-E}$), while the difference in runoff ($\Delta\,F_R$) is negligible. A closer look (Figure~\ref{fig:supPmE}\textbf{A})  reveals that this switch in the sign of the $P-E$ flux difference occurs because both the difference in rain and snow, as well as in evaporation switch sign. These in turn are mainly related to the reversal in the difference in sea surface temperature (SST) in the North Atlantic, which in turn is due to the reversal in the difference of northward heat transport by the AMOC. In other words, the behaviour of the $P-E$ flux is a result of the heat transported by the AMOC, and hence a self-reinforcing mechanism: the stronger AMOC sees a net smaller $F_{P-E}$ flux. Moreover, under CBS the southward sea ice export from the Arctic is smaller for low hosing, and then this difference also switches sign for increased hosing (Figure~\ref{fig:supPmE}\textbf{C}). As a consequence we find the same pattern in the total sea ice area in the North Atlantic, where a larger sea-ice area limits evaporation from the ocean's surface. 

To summarize the effect of the Bering Strait  closure on the AMOC ON state,  it  affects the Atlantic-Arctic exchange in two ways. For low hosing values --i.e. a strong AMOC-- the southward freshwater and sea-ice exports are smaller  for a closed Strait than for an open Strait. As a result the North Atlantic is much more saline (Figure~\ref{fig:supSSS}\textbf{A}). With the resulting stronger AMOC we also find a larger total freshwater import, increased SSTs in the North Atlantic, and increased precipitation and evaporation over the North Atlantic. For increased hosing --and so a weaker AMOC-- we no longer see a marked difference in the lateral freshwater transports between OBS and CBS, but the sea-ice import and area in the North Atlantic is now larger under CBS. This limits evaporation, and weakens the AMOC, resulting in lower SSTs and a larger surface freshwater flux. This means the North Atlantic is now fresher under CBS (Figure~\ref{fig:supSSS}\textbf{B}) and explains the lower critical hosing value for AMOC 
tipping. Furthermore, there seems to be a critical hosing level $F_{H,c}\approx 0.161$~Sv, beyond which an open Bering Strait has a salinifying instead of a freshening effect on the North Atlantic. This corresponds to an equilibrium AMOC strength dropping beneath $16.4$~Sv, equivalent to a decrease of $16.3\%$ from pre-industrial strength.

\subsubsection*{CO$_2$-forcing}
Next, the equilibria of an AMOC ON-state under an open Bering Strait for various fixed hosing values $F_H\in[0.00, 0.15]$~Sv are taken, and forced by a $1\%$ CO$_2$ increase per year (starting at $280$~ppm) until a prescribed amount of carbon emissions is reached after which the emission rate is set to zero, following the ZECMIP protocol~\cite{jones2019zero}. These experiments  are repeated, but now the Bering Strait is immediately closed at the start of the simulation. For  both OBS and CBS cases,  the safe carbon budget --i.e. the maximum amount of carbon emissions without an AMOC collapse-- is determined to within $100$~PgC. 

\begin{figure}
    \centering
    \includegraphics[width = \textwidth]{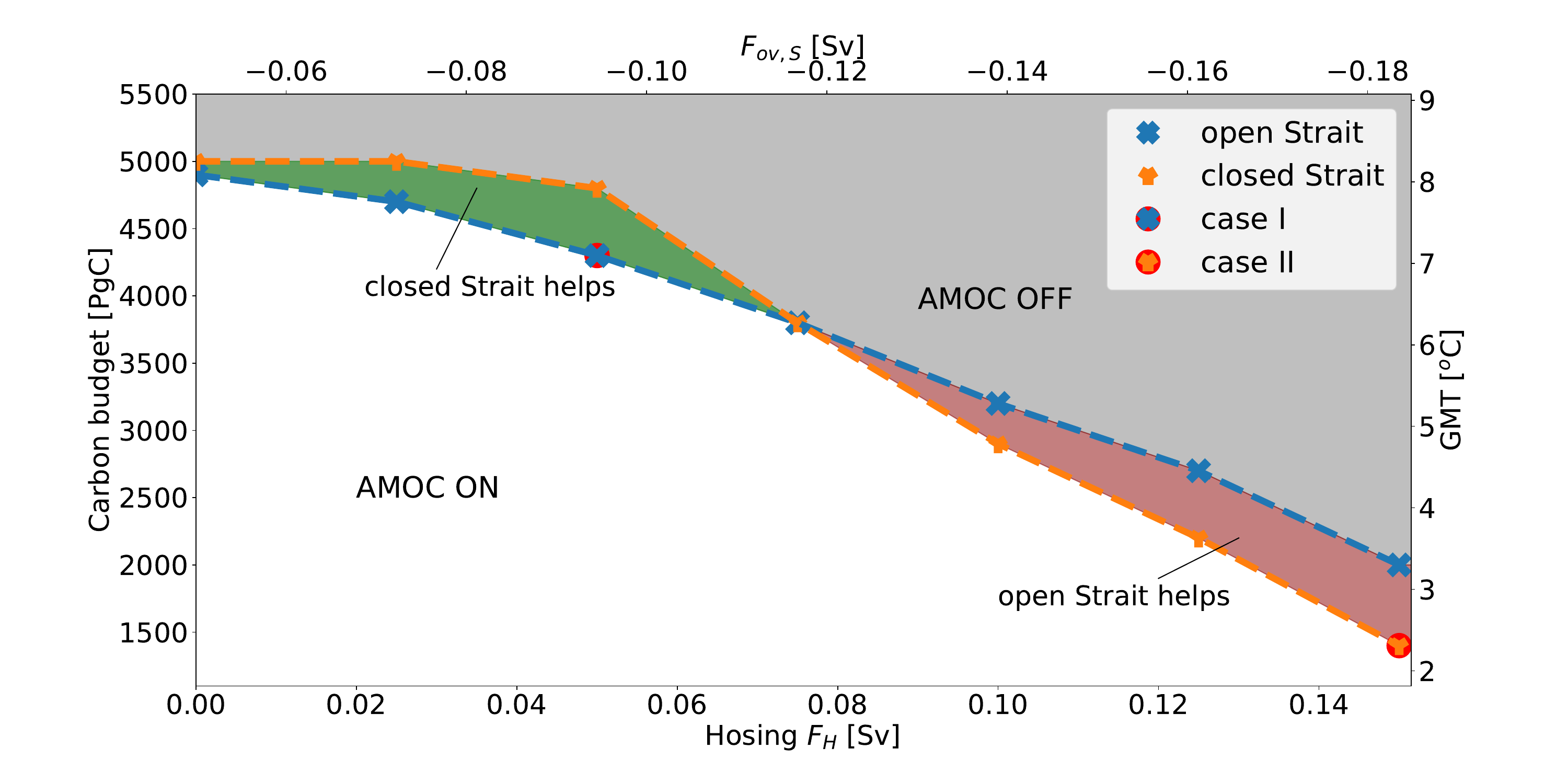}
    \caption{\textbf{$1$~$\%$/yr  CO$_2$-forcing experiment.} The safe carbon budget of the AMOC under OBS and CBS with a starting AMOC state under OBS at various fixed hosing values $F_H$. The CO$_2$-forcing is increased at a $1\%$ rate until the budget is reached, where either the Bering Strait is kept open (blue marks), or directly closed at the start of the simulation (orange marks). Case I and II are indicated with an additional red dot. The gray (white) region indicates hosing and carbon budget values under which the AMOC collapses (does not collapse) regardless whether the Strait is closed or open. The green (red) region indicates forcing values under which the AMOC only collapses if the Strait is open (closed). On the right axis the global mean temperature (GMT) increase corresponding to the carbon budget emitted is indicated, using $1.65\,^\circ$C/$1000$~PgC~\cite{ipcc2021wg1}, and on the top axis the approximate corresponding $F_{ov,S}$ value of the starting equilibrium state using the linear fit in Figure~\ref{fig:supfit}.}
    \label{fig:1percent}
\end{figure}

The results are shown in Figure~\ref{fig:1percent}. The white and gray regions indicate the amount of carbon emissions and fixed hosing values under which a direct closure of the Bering Strait does not alter the outcome. For hosing values $F_H \in [0.00, 0.075]$~Sv the safe carbon budgets are higher when the Strait is directly closed at the start of the simulation, while for higher hosing values the reverse is true and an immediate closure of the Strait would actually reduce the safe carbon budget. The reported emission amounts are related to the corresponding rise in global mean temperature (GMT), using that approximately $1.65\,^\circ$C/$1000$~PgC~\cite{ipcc2021wg1}. Moreover, the hosing values are related to the freshwater transport induced by the overturning circulation in the Atlantic at $35^\circ$S ($F_{ov,S}$) in equilibrium, using a least-squares linear fit, see Figure~\ref{fig:supfit}. This is done as $F_{ov,S}$ is an important indicator for the strength of the AMOC's salt-advection feedback and can reveal any biases in the Atlantic's freshwater budget~\cite{weijer2019stability,van2024persistent, van2024physics, vanderborght2024feedback}. In this model the computed $F_{ov,S}$ lie neatly within the observed range~\cite{garzoli2013south}.

\begin{figure}
    \centering
    \includegraphics[width = \textwidth]{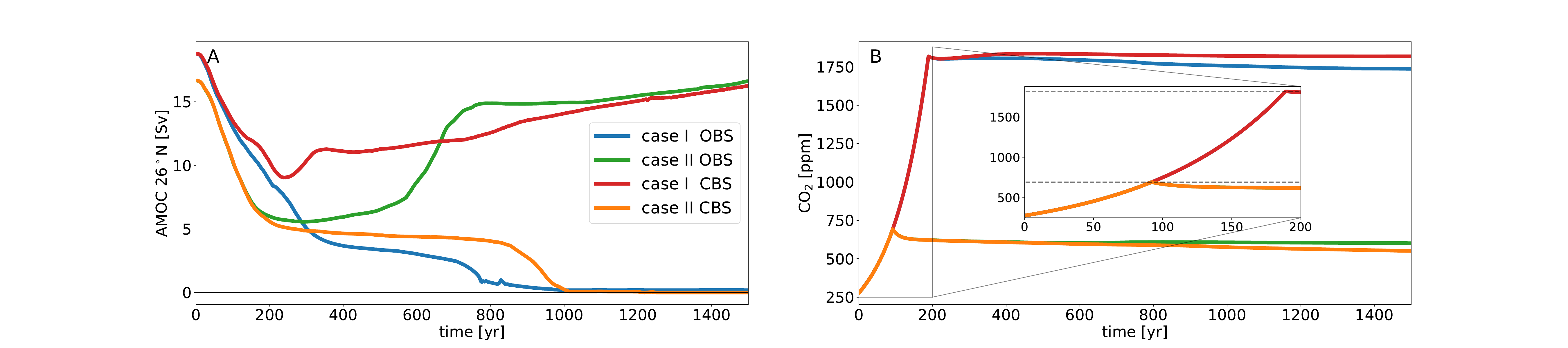}
    \caption{\textbf{Case I \& II.} Case I with a $1$~$\%$/yr CO$_2$ increase for $188$~yr and hosing $F_H = 0.05$~Sv with an open Strait (blue) and an immediate closure (red), and Case II with a $1$~$\%$/yr CO$_2$ increase for $93$~yr and hosing $F_H = 0.15$~Sv with an open Strait (green) and an immediate closure (orange), showing the AMOC strength (\textbf{A}), and corresponding atmospheric CO$_2$-concentration (\textbf{B}). The horizontal dashed lines indicate the maximum attained CO$_2$-concentration, which is $1820$~ppm and $693$~ppm for case I and II respectively. Note that the CO$_2$-concentration drops faster if the AMOC has collapsed, since this affects the marine carbon uptake~\cite{boot2024response}.}
    \label{fig:cases}
\end{figure}

We will treat two forcing scenarios in more detail (Figure~\ref{fig:1percent}). We consider case I, where the forcing consists of a hosing value of $0.05$~Sv with a $1$~$\%$/yr increase in CO$_2$ for $188$~yr before emissions are set to zero (leading to a total of $4300$~PgC of emissions), and case II with a hosing value of $0.15$~Sv with a $1$~$\%$/yr increase in CO$_2$ for $93$~yr before emissions are set to zero (leading to a total of $1400$~PgC of emissions). In both cases the initial CO$_2$ concentration is $280$~ppm. In case I,  an immediate closure of the Bering Strait prevents an AMOC collapse, while in case II  a collapse occurs only because of the closure (Figure~\ref{fig:cases}). For all 
trajectories the AMOC shows a steep drop in strength under the CO$_2$-forcing, with a weakening up to $8.5$~Sv within the first $300$~yr. The corresponding CO$_2$-concentrations are quite extreme: in case II the maximum concentration of $693$~ppm is reached within a century, and in case I the maximum of $1820$~ppm is attained within two centuries. Under the most extreme shared socio-economic pathway SSP5-8.5 a global average concentration of $1135.2$~ppm in 2100~CE and $2108.3$~ppm in 2200~CE is reached~\cite{meinshausen2020shared}. Within these first $300$~yr the recovering trajectories (case I under CBS, and case II under OBS) reach their AMOC minima (at $241$~yr and $292$~yr, respectively), and therefore the key factor determining whether the AMOC collapses or recovers must already be present in the 
first three centuries. 

\begin{figure}
    \centering
    \includegraphics[width = \textwidth]{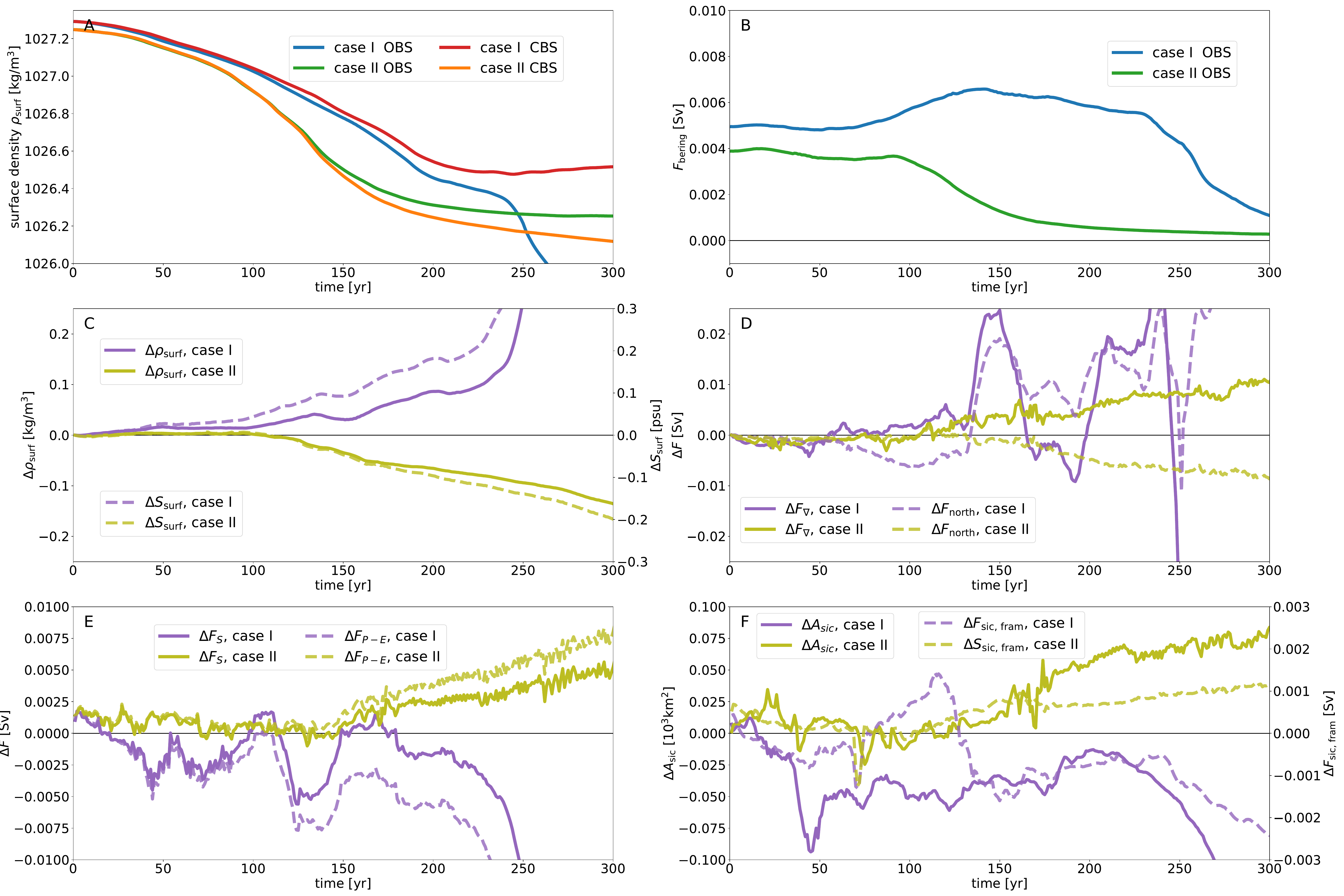}
    \caption{\textbf{Case I \& II diagnostics.} Case I with a $1$~$\%$/yr CO$_2$ increase for $188$~yr and hosing $F_H = 0.05$~Sv with an open Strait (blue) and an immediate closure (red), and Case II with a $1$~$\%$/yr CO$_2$ increase for $93$~yr and hosing $F_H = 0.15$~Sv with an open Strait (green) and an immediate closure (orange) with their average density $\rho_{\text{surf}}$ of the top $200$~m of the North Atlantic region (\textbf{A}), and the freshwater transport through the Bering Strait $F_{\text{bering}}$ (\textbf{B}). Moreover, the difference between CBS and OBS settings for case I (purple) and case II (citrus) in surface density $\Delta\rho_{\text{surf}}$ (\textbf{C}, solid) and in surface salinity $\Delta S_{\text{surf}}$ (\textbf{C}, dashed), in freshwater import through the lateral boundaries $\Delta F_{\nabla}$ (\textbf{D}, solid) and in freshwater import through then northern boundary $\Delta F_{\text{north}}$ (\textbf{D}, dashed), in surface freshwater transport $\Delta F_S$ (\textbf{E}, solid) and in precipitation-minus-evaporation $\Delta F_{P-E}$ (\textbf{E}, dashed), and in sea-ice area in the North Atlantic $\Delta A_{\text{sic}}$ (\textbf{F}, solid) and in southward sea-ice export through the Fram Strait $\Delta F_{\text{sic, fram}}$ (\textbf{F}, dashed).}
    \label{fig:casediag}
\end{figure}

In Figure~\ref{fig:casediag} the surface densities, and differences in freshwater fluxes into the North Atlantic region for case I and II 
are depicted during the first $300$~yr of the simulation simulations. Figure~\ref{fig:supcasediag} depicts the same quantities over 
the full simulation runtime. Figures~\ref{fig:casediag}\textbf{A\&C} show the average surface density in the North Atlantic, and the difference in surface density and salinity between the CBS and OBS settings, respectively. Already at the start of the simulations,  the density of the collapsing trajectory is lagging behind. Moreover, this coincides mainly with a lag in salinity, and therefore the cause must be the freshwater inputs into the North Atlantic region. Considering the difference in freshwater import through its lateral boundaries ($\Delta F_{\nabla}$, Figure~\ref{fig:casediag}\textbf{D}) we see that the signal for case II is quite consistent. Here for most of the first $300$~years there is a higher freshwater import under a closure of the Strait, despite that the freshwater import from the Arctic is lower, i.e. $\Delta F_{\text{north}}<0$. 

For case II we also see a consistently higher surface freshwater forcing under a closure ($\Delta F_S > 0$, see Figure~\ref{fig:casediag}\textbf{E}), mainly due to a higher precipitation-minus-evaporation over the North Atlantic. Again we can relate this to a higher sea-ice coverage --as this limits evaporation-- over the North Atlantic, which is the result of a higher sea-ice export from the Arctic to the North Atlantic, see Figure~\ref{fig:casediag}\textbf{F}. Hence, the lowered surface density and salinity under a closure for case II is despite the lowered freshwater import from the Arctic, and mainly due to the increased $F_{P-E}$ flux over the North Atlantic, which is related to the increased sea-ice export from the Arctic. Note also that for case II the freshwater import from the Pacific into the Arctic is actually elevated during the first $250$~yr despite the weakened AMOC. 

Understanding the behavior in case I, where a closure prevents an AMOC collapse, is  less straightforward. The difference in the total import of freshwater through the lateral boundaries $\Delta F_{\nabla}$ is quite erratic, but mainly higher under CBS for the first $250$~years before dropping steeply. The freshwater import from the Arctic on the other hand is initially lower under CBS settings for the first $130$~years (Figure~\ref{fig:casediag}\textbf{D}). Hence the initial higher surface density and salinity under a closure is partly due to the reduced freshwater import from the Arctic, but only for roughly the first century of the simulation.  For the full simulation, the surface freshwater flux  is lower under a closure, which is mainly due to a lower precipitation-minus-evaporation (Figure~\ref{fig:casediag}\textbf{E} and Figure~\ref{fig:supcasediag}\textbf{E}). This --again-- can be related to the lower sea-ice coverage in the North Atlantic with the  reduced sea-ice export from the Arctic. Hence overall the more saline North Atlantic under CBS can be explained by the lower $F_{P-E}$  due to the lower sea-ice cover, and partly by the reduced freshwater import from the Arctic. Note that for case II we see that $F_{\text{bering}}$ reduces more rapidly than in case I, and hence under OBS less freshwater is imported to the Arctic from the Pacific in case II than in case I. This agrees with a closure preventing an AMOC collapse in case I but not in case II. 

The results of the CO$_2$-forcing experiment align partly with those of the hysteresis experiment. For a low hosing value (e.g. $F_H = 0.05$~Sv) a closure results in a more saline North Atlantic, and hence a more resilient AMOC, as it limits the freshwater surface forcing via reduced sea-ice export from the Arctic and it limits the freshwater export from the Arctic. For a high hosing value (e.g. $F_H = 0.15$~Sv) a closure still reduces the freshwater import from the Arctic, but the increased surface forcing over the North Atlantic --as we also saw for the hysteresis experiment-- causes a fresher Atlantic and hence a more vulnerable AMOC. Note that the freshwater import through the Bering Strait under CO$_2$-forcing does not align with the results for the freshwater-forced hysteresis experiment, as has also been seen in a previous study~\cite{hu2023dichotomy}. Despite the AMOC weakening under CO$_2$-forcing in case I this freshwater import actually increases initially. Also here it seems that there is a critical hosing value $F_{H, c}$ above which a closure has a net freshening instead of salinifying effect on the North Atlantic. In case of the CO$_2$-forcing employed here we found $F_{H, c} = (0.075\pm 0.0125)$~Sv, which corresponds to an equilibrium AMOC strength of $(18.4\pm0.2)$~Sv under OBS. This is equivalent to a reduction in strength of $(6.1\pm0.5)\%$ from pre-industrial. We expect that this $F_{H,c}$ value is dependent on the type and rate of the applied forcing.

\subsubsection*{Delay of the Closure}
To study the effect of the timing of the Bering Strait closure, we take the hosing values  for which a closure extends the AMOC's safe carbon budget, and we apply again a $1$~$\%$/yr CO$_2$-forcing starting at $280$~ppm. This entails the forcing scenarios $F_H = 0.0$~Sv with a CO$_2$ increase over $202$~yr, $F_H = 0.025$~Sv with a CO$_2$ increase over $198$~yr, and $F_H = 0.05$~Sv with a CO$_2$ increase over $188$~yr. This results in a total amount of emissions of $4900$~PgC, $4700$~PgC and $4300$~PgC, respectively. They all fall within the previously determined extended safe budget for an immediate closure. However, we now delay a closure of the Bering Strait by either $50$~yr, $100$~yr, $150$~yr, $200$~yr, $250$~yr, or $300$~yr after the forcing has started. 

\begin{figure}
    \centering
    \includegraphics[scale = 0.25]{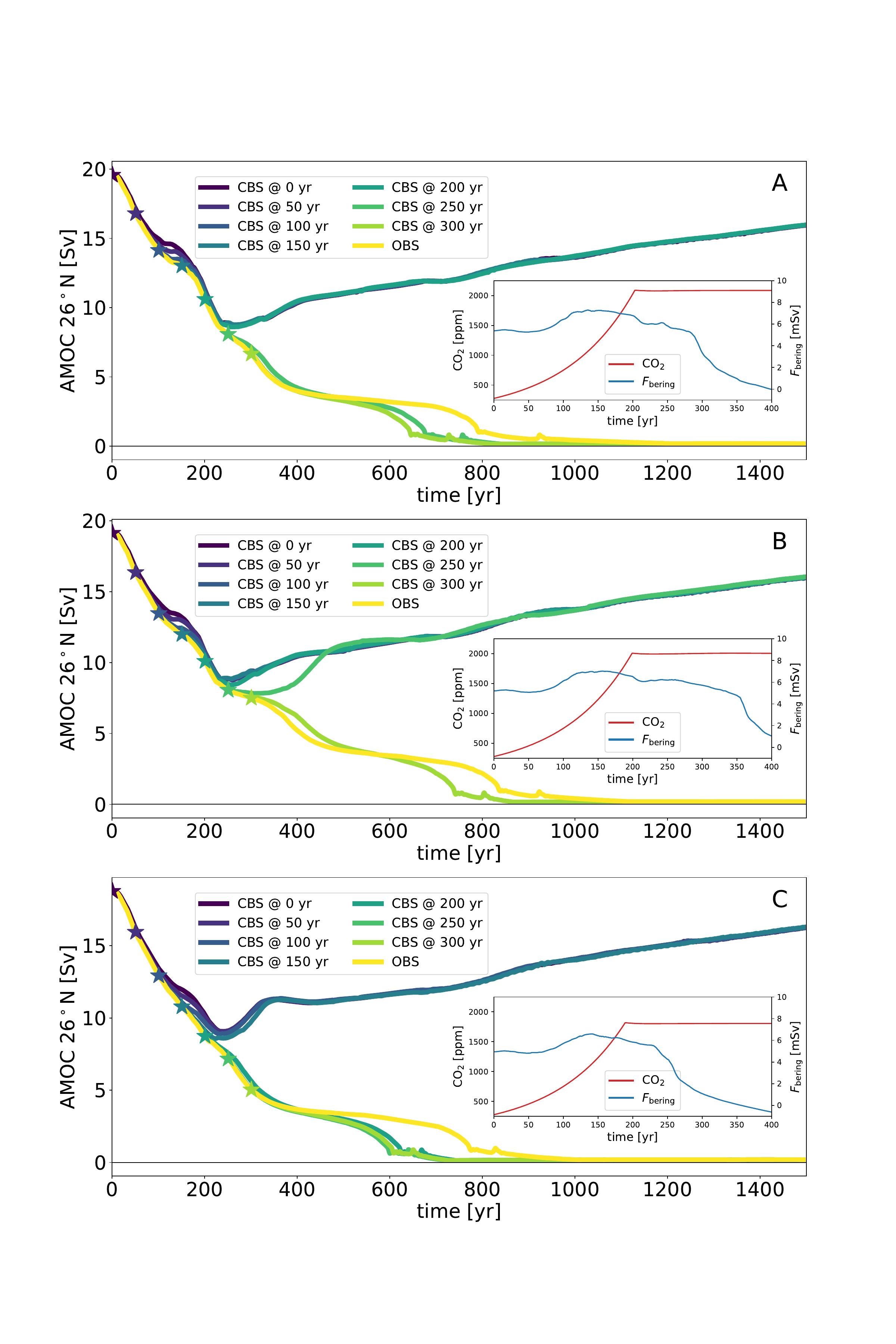}
    \caption{\textbf{Delay of the closure.} The AMOC strength under three forcing scenarios for which an immediate closure of the Bering Strait prevents an AMOC collapse: $F_H = 0.0$~Sv with $4900$~PgC of emissions (\textbf{A}), $F_H = 0.025$~Sv with $4700$~PgC of emissions (\textbf{B}), and $F_H = 0.05$~Sv with $4300$~PgC of emissions (\textbf{C}). The closure is done either at time $0$~yr, $50$~yr, $100$~yr, $150$~yr, $200$~yr, $250$~yr, $300$~yr, or not at all (OBS) (purple to yellow, solid). The asterisks in corresponding colors indicate time of closure. The insets show the corresponding CO$_2$-concentrations (red, left axis) and freshwater flux through the Strait (blue, right axis) in the first $400$~yr for the open Strait scenario.}
    \label{fig:timing}
\end{figure}

The closure of the Bering Strait can be delayed up to $200$~yr, $250$~yr, and $150$~yr (Figure~\ref{fig:timing}) for the three forcing scenarios, respectively. A closure that is too late to prevent a collapse, actually speeds up the AMOC's collapse. This is again in agreement with the dynamical results from the hysteresis experiment. For low hosing values and a sufficiently strong AMOC a closure inhibits the Arctic's freshwater transport to the North Atlantic, and hence strengthens the AMOC. However, if the AMOC is weakened or OFF then for an open Bering Strait there is a reduction of the freshwater fluxes into the North Atlantic, and so a closure would weaken the AMOC. For the scenarios presented in Figure~\ref{fig:timing}\textbf{A-C} the critical AMOC strength below which a closure causes a weakening is then $9.4\pm1.3$~Sv, $7.8\pm0.3$~Sv and $9.8\pm1.0$~Sv, respectively. Note that for the second scenario a preventive closure is still effective for a relatively low AMOC strength. In this scenario the freshwater transport through the Bering Strait is relatively high for these low strengths, and so an OBS still has a freshening effect on the Arctic. That this differs with the other scenarios might be related to the fact that the safe carbon budget is determined in increments of $100$~PgC, and so in the second scenario the forcing can be closer to the actual safe carbon budget, and consequently  a lower AMOC is reached without tipping. It is encouraging that there is a large window of opportunity to construct the BSD despite a significant AMOC 
weakening having already taken place.

\subsection*{Discussion}

The results presented here indicate that an artificial closure of the Bering Strait can be an effective climate intervention strategy in order to prevent an AMOC collapse under CO$_2$-forcing. The hysteresis experiment showed the dynamical effect of a closure on the AMOC, and agreed qualitatively with existing literature \cite{hu2005bering, hu2012role, hu2015effects, hu2023dichotomy, cessi2020control, otto2017amplified, de2004bering, weiffenbach2025impact}. For an active AMOC with a relatively saline North Atlantic a closure will allow for a reduced freshwater transport out off the Arctic into the North Atlantic, and hence cause an AMOC strengthening. For an increasingly fresher North Atlantic and weaker AMOC an open Bering Strait has a stabilizing effect, as it allows for more freshwater to leave the North Atlantic and reduced sea-ice import into it. This effect is even more pronounced for a collapsed AMOC. The CO$_2$-forcing experiment again showed that a relatively strong AMOC sees a net freshening effect of the Bering Strait, and so its closure can prevent a CO$_2$-induced AMOC tipping, while for a weaker AMOC in the initial equilibrium (i.e. increased hosing value $F_H$ beyond the critical hosing $F_{H,c}$) the reverse is true. Lastly, the closure delay experiment demonstrated that a closure occurring when the AMOC is already severely weakened has a counterproductive effect. However, it also showed an operating window of at least $150$~yr after the $1\%$ CO$_2$-forcing had commenced. 

The technical feasibility of the BSD is supported by the fact that its construction challenges are on par with already completed mega-projects. As stated earlier, the BSD would have a total length of roughly $80$~km, with an average and maximum depth of $50$~m and $59$~m, respectively. By way of comparison, the current largest enclosure dam is the Saemangeum Seawall (South Korea) with a length of $33$~km and a maximum depth of $54$~m~\cite{groeskamp2020need}. So in both dimensions the BSD would be of the same order of magnitude. Moreover, if we assume the BSD to rise $20$~m above sea-level and to be $100$~m wide at the top with two sloping sides of $1:2$ (height : width) ratio~\cite{jonkman2013costs}, then roughly $1.3$ km$^3$ of raw material is needed to build the dam. This is only a factor $3.5$ larger than the amount needed to construct Maasvlakte 2, the extension of the Port of Rotterdam~\cite{groeskamp2020need}. Hence, also in this aspect the required dimensions stay within the same order of magnitude as already developed projects. A more detailed feasibility analysis is beyond the scope of this study.

Although CLIMBER-X is ideally suited to study the detailed mechanisms of the effects of Bering Strait closure on AMOC stability
under climate change, its horizontal spatial resolution of $5^\circ\times 5^\circ$ is very coarse. In fact, the BST is modeled as a 
baroclinic tracer exchange between the Arctic and North Pacific. Hence the model does only provide the freshwater and heat 
exchange through the Strait. As already mentioned above, this gives discrepancies with observations on volume and 
freshwater transport values \cite{haine2015arctic}. However, qualitatively the dynamics in CLIMBER-X coincide with existing 
studies, some  done at higher resolutions \cite{de2004bering, hu2005bering, hu2012role, hu2015effects}, and so we are 
confident in the  results and, more important, the generality of the physical mechanisms. 

Another limitation is in the choice of the applied forcings. This hosing was 
varied in order to explore the response of various AMOC ON states with different freshwater budgets and strengths. This is done as the 
freshwater budget of the pre-industrial AMOC has slight biases in the model~\cite{willeit2022earth, van2024persistent}. 
As a consequence however the freshwater hosing does become unrealistically high up to $0.15$~Sv, which is roughly 
a factor $20$ larger than the present-day melt rate of the Greenland Ice Sheet~\cite{sasgen2020return}. 
The climate forcing consisted of solely the CO$_2$-concentration being increased at a $1\%$ rate per year. This was 
done instead of a SSP scenario for the sake of simplicity albeit less realistic~\cite{pielke2022plausible}. In order 
to cause an AMOC collapse
total carbon emissions of up to $5000$~PgC were needed. Although large, this amount still falls below the upper estimate of available fossil fuel reserves~\cite{ipcc2001}, and the corresponding CO$_2$-concentrations fall below those computed for the extended SSP5-8.5 scenario as well~\cite{meinshausen2020shared}. 

To get a more quantitative assessment on whether a closure of the Bering Strait is able to prevent an AMOC collapse in our present-day climate, a closure experiment needs to be done in a more detailed climate model --including more valid Arctic freshwater transports--  under more realistic climate forcings. This will allow for a more accurate quantitative assessment of the BSD's effect. A next step is therefore to simulate a closure in the Community Earth System Model (CESM): a state-of-the-art climate model in which an AMOC collapse has been found under SSP-4.5 forcing scenario with a more accurate Atlantic freshwater budget~\cite{van2025european, van2025phys}. As it is qualitatively now expected that there is a critical hosing value $F_{H,c}$ (and corresponding equilibrium AMOC strength) below (above) which a closure aides the AMOC's resilience to this climate forcing, additional simulations are needed to improve the estimate of this value. Lastly, for a full consideration of the BSD as an alternative climate intervention analyses regarding its technical feasibility, economic effects, and environmental impacts are needed. We expect the BSD to have a large impact onto the regional ecosystem~\cite{council2009arctic}, and so especially in this regard we do want to stress that CO$_2$ mitigation efforts 
are the preferable option to avoid an AMOC collapse. But if this is not realized, this study showed that in an EMIC a man-made 
timely closure of the Bering Strait can prevent a collapse of the AMOC under climate change. 

\newpage

\begin{figure}
    \centering
    \includegraphics[width = \textwidth]{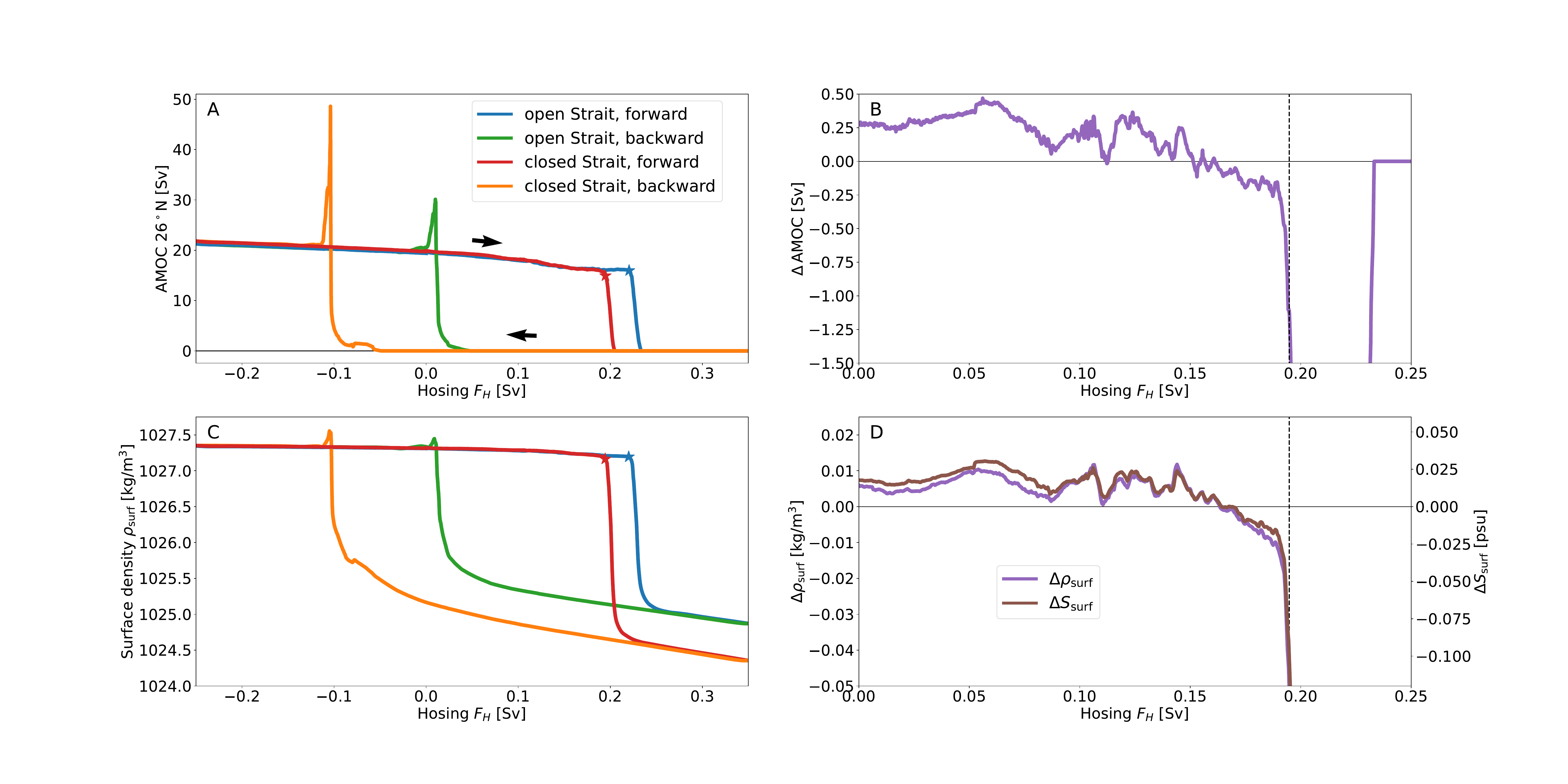}
    \caption{\textbf{The hysteresis experiment.} The quasi-equilibrium simulations for an open Strait (blue, green), and a closed Strait (red, orange), consisting of simulations where the hosing flux $F_H$ increases (blue, red), and decreases (green, orange). The asterisks mark the estimated tipping points of the AMOC collapses (\textbf{A}\&\textbf{C}), while the vertical line (dashed, black) indicates the critical hosing value for the AMOC tipping under CBS (\textbf{B}\&\textbf{D}. (\textbf{A}) The AMOC strength --computed as the maximum overturning strength at $26^\circ$N-- for varying hosing flux $F_H$, where the arrows indicate the direction of time during the hosing experiment. The difference in AMOC strength ($\Delta\,$AMOC) between the ON-states under CBS and OBS is shown in (\textbf{B}). (\textbf{C}) The average density $\rho_{surf}$ of the top $200$~m surface layer of the North Atlantic region between $50^\circ$N and $75^\circ$N, and the difference in average density $\Delta\rho_{\text{surf}}$ (purple) and average salinity $\Delta S_{\text{surf}}$ (brown) in this layer between the ON-states under CBS and OBS (\textbf{D}).}
    \label{fig:hysteresis}
\end{figure}

\clearpage 

%
\bibliography{references} 
\bibliographystyle{sciencemag}

%
%
%
%
%
%


\paragraph*{Acknowledgments:}
The authors would like to thank Amber Boot and Michael Denes for their help with running the simulations.

\paragraph*{Funding:}
J.S. and H.A.D. are funded by the European Research Council through ERC-AdG project TAOC (project 101055096).

\paragraph*{Author contributions:}
J.S. conceived the idea for this study, performed the model simulations, conducted the first analyses and prepared the figures. H.A.D. acquired the funding. Both authors were actively involved in the interpretation of the analysis results and the writing process.

\paragraph*{Competing interests:}
There are no competing interests to declare.

\paragraph*{Data and materials availability:}
The (processed) model output and analysis scripts are provided at \url{https://doi.org/10.5281/zenodo.16949317} and \url{https://doi.org/10.5281/zenodo.16949298}.


\subsection*{Supplementary materials}
Materials and Methods\\
Freshwater forcing for backward hysteresis simulations\\
Figs. S1 to S7


\newpage


\renewcommand{\thefigure}{S\arabic{figure}}
\renewcommand{\thetable}{S\arabic{table}}
\renewcommand{\theequation}{S\arabic{equation}}
\renewcommand{\thepage}{S\arabic{page}}
\setcounter{figure}{0}
\setcounter{table}{0}
\setcounter{equation}{0}
\setcounter{page}{1} 


\begin{center}
\section*{Supplementary Materials for\\ \scititle}

Jelle~Soons$^{\ast}$,
Henk~A.~Dijkstra\\ 
\small$^\ast$Corresponding author. Email: j.soons@uu.nl
\end{center}

\subsubsection*{This PDF file includes:}
Materials and Methods\\
Freshwater forcing for backward hysteresis simulations\\
Figures S1 to S7

\newpage


\subsection*{Materials and Methods}
\subsubsection*{CLIMBER-X}
CLIMBER-X~\cite{willeit2022earth, willeit2023earth} is a fast Earth system Model of Intermediate Complexity (EMIC) employing the frictional geostrophic 3D ocean model GOLDSTEIN~\cite{edwards2005uncertainties} together with the semi-empirical statistical-dynamical atmospheric model SESAM~\cite{willeit2022earth}, a dynamic-thermodynamic sea ice model SISIM~\cite{willeit2022earth}, and the land surface model with interactive vegetation PALADYN~\cite{willeit2016paladyn}. There are also components for ocean biogeochemistry (HAMOCC) and ice sheets (SICOPOLIS or Yelmo), but there are not used in this study. Note that therefore ice sheets are prescribed at their modern state, and so the net freshwater flux from these sheets is assumed to be zero. Hence, in our simulations with climate forcing the AMOC is not affected by increased melt from e.g. the Greenland Ice Sheet. As mentioned before, all components of the climate model have a horizontal resolution of $5^\circ\times 5^\circ$. 

The atmosphere model SESAM (Semi-Empirical dynamical Statistical Atmosphere Model) uses a combination of observational data as well as results from Global Climate Models (GCMs) where all prognostic variables (e.g. temperature, humidity, wind speed) are determined on a 2D grid, while the vertical structure is purely diagnostic. The general vertical structure in the atmosphere of humidity and temperature are used to determine the complete 3D structure of these variables, while the thermal wind balance is employed to compute the 3D structure of the wind. Longwave radiation fluxes take into account several greenhouse gases such as methane, CFCs, ozone and CO$_2$. Clouds are also represented with one cloud layer which is characterized by variables such as cloud fraction and albedo. 

The ocean model GOLDSTEIN is run on 23 non-equidistant vertical layers. Horizontal velocities are derived from a frictional-geostrophic balance, while vertical velocities follow from the continuity equation. Throughout the water column a hydrostatic balance is assumed. Moreover, a rigid-lid approximation is assumed, and therefore surface freshwater fluxes are represented as virtual salinity fluxes. Note that the $5^\circ\times 5^\circ$ rectilinear grid is too coarse to represent the Bering Strait directly: the Strait is enclosed in the grid cell centered at $67.5^\circ$N and $167.5^\circ$W, which has an ocean fraction of $0.84$. An open Strait is modeled by allowing baroclinic tracer exchange between the Arctic and North Pacific Ocean, while the Strait is always closed for barotropic flow. A closure of the Strait entails a seizing of the tracer exchange. Consequently, the Throughflow's strength in this model is not realistic, while its effect on the buoyancy of the North Atlantic is, as the exchange of tracers such as freshwater and temperature is captured. The model's performance is comparable with state-of-the-art CMIP6 models under various forcings and boundary conditions. In particular the deep-convection zones in the model coincide with those following ocean reanalysis, while the AMOC's overturning pattern at $26^\circ$N is quite similar to the RAPID observations, although the modeled AMOC is a bit too shallow. Moreover, due to the strong momentum damping the Antarctic Circumpolar Current is too weak.

The sea-ice model SISIM (Simple Sea Ice Model) models sea-ice as one snow layer on top of one ice layer. The snow can accumulate and melt, and if it exceeds $1$~m then the excess becomes ice. This ice layer cannot only accumulate from above, but also grow via accretion from below, and melt from above and from below. The freezing temperature is dependent on the local ocean salinity. Moreover, the sea-ice is allowed to drift. Lastly, SISIM also acts as an ocean-atmosphere coupler even in sea-ice free regions.

The land module PALADYN computes the water and energy fluxes between the land surface, soil and atmosphere. It represents the terrestrial carbon cycle including dynamical vegetation. Water, temperature and carbon contents are solved in the soil using five vertical layers.

\subsubsection*{Climate model simulations}
The hysteresis experiment is performed with a prescribed freshwater flux $F_H$ into the Atlantic latitudinal belt $20^\circ$N to $50^\circ$N. This freshwater hosing is compensated globally. The first hysteresis simulation starts in the pre-industrial equilibrium, i.e. no hosing and with CO$_2$ at 280 ppm, after which the hosing is increased at a rate of $0.025$~Sv/kyr till the total hosing is $0.35$~Sv. Then it is reduced with a rate of $-0.025$~Sv/kyr till the total hosing is $-0.25$~Sv, and finally it is again increased with the same rate back to zero hosing. The total simulation takes $48000$~yr. For the second hysteresis we repeat this protocol, but we start in the equilibrium with a closed Bering Strait and active AMOC under pre-industrial conditions. This state is computed with a $7000$~yr long simulation that starts in the original pre-industrial equilibrium where the Strait is directly closed at start.

For the $1\%$ CO$_2$-forcing experiment the starting states for fixed freshwater fluxes with an open Strait are taken from the hysteresis simulation. The CO$_2$-concentration is increased at $1\%$/yr (starting at $280$~ppm) until the prescribed emission budget is reached after which no emissions occur, following the ZECMIP protocol~\cite{jones2019zero}. The simulations run for $1500$~yr. At the start of the simulation the Strait is either closed or kept open. The safe budget is found by increasing the emissions with increments of $100$~PgC until the AMOC collapses. 

For the third experiment this set-up is repeated for the mentioned forcing scenarios, but now the closure is delayed up to $300$~yr in increments of $50$~yr after the simulation start.

\subsubsection*{The AMOC strength}
The AMOC strength is defined as the maximum of the Atlantic meridional overturning streamfunction $\psi_A(y,z)$ at $26^\circ$N
\begin{equation*}
    \text{AMOC} = \text{max}_{y = 26^\circ\text{N}}\left[\psi_A(y,z)\right]
\end{equation*}
where the streamfunction is computed as
\begin{equation*}
    \psi_A(y,z) = -\int_{z}^0\int_{x_W(y,z)}^{x_E(y,z)}\,v(x',y,z')\,dx'\,dz'
\end{equation*}
where $v$ is the meridional velocity, and $x_W$ and $x_E$ are the western and eastern boundary of the Atlantic basin, respectively. 

\subsubsection*{AMOC tipping point estimate}
The tipping points of an AMOC collapse based on the hysteresis simulations in figure~\ref{fig:hysteresis} are determined as the last point before the collapse where it still holds that $\frac{\partial\text{AMOC}}{\partial F_H} > -1$. During an AMOC collapse we have $\frac{\partial\text{AMOC}}{\partial F_H} < -1$: the changes in AMOC strength are then primarily driven by internal feedbacks instead of changes in the external forcing.

\subsubsection*{Freshwater transports into the North Atlantic region}
The region we consider the North Atlantic is the region in the Atlantic between the latitudes $50^\circ$N and $75^\circ$N and longitudes $75^\circ$W and $55^\circ$E, see figure~\ref{fig:supregion}. This region is chosen as it encompasses most of the mixed layer zones in the model. The surface density (or salinity) is computed as the average density (or salinity) of the top $200$~m layer in this region. We compute four freshwater transports, one for each boundary section. A positive sign indicates a net import into this region. These are computed as
\begin{align*}
    F_{\text{north}} &= -\int_{-H}^0\int_{75^\circ\text{W}}^{55^\circ\text{E}}\, v \left(1-\frac{S}{S_0}\right)\Bigg|_{75^\circ\text{N}}\,dx\,dz\\
    F_{\text{east}} &= -\int_{-H}^0\int_{50^\circ\text{N}}^{75^\circ\text{N}}\, u \left(1-\frac{S}{S_0}\right)\Bigg|_{55^\circ\text{E}}\,dy\,dz\\
    F_{\text{south}} &= \int_{-H}^0\int_{75^\circ\text{W}}^{55^\circ\text{E}}\, v \left(1-\frac{S}{S_0}\right)\Bigg|_{50^\circ\text{N}}\,dx\,dz\\
    F_{\text{west}} &= \int_{-H}^0\int_{50^\circ\text{N}}^{75^\circ\text{N}}\, u \left(1-\frac{S}{S_0}\right)\Bigg|_{75^\circ\text{W}}\,dy\,dz
\end{align*}
where $u$ and $v$ are the zonal and meridional velocity, respectively, $S_0 = 34.7$~psu is the reference salinity, and $H$ the local water depth. Then the net freshwater import through the zonal boundaries of the North Atlantic ($F_{\text{zonal}}$), and the net import through all lateral boundaries ($F_{\nabla}$) can be computed.
\begin{align*}
    F_{\text{zonal}} &= F_{\text{east}} + F_{\text{west}}\\
    F_{\nabla} &= F_{\text{north}} + F_{\text{east}} + F_{\text{south}} + F_{\text{west}}
\end{align*}
The freshwater transport through the Bering Strait is similarly computed as the meridional freshwater transport through the 65th parallel north between the Pacific and Arctic Ocean. 

We also consider the freshwater flux $F_S$ through the North Atlantic's surface, following
\begin{equation*}
    F_{S} = F_H + F_R + F_{P-E} + F_{\text{rest}}
\end{equation*}
where $F_H$ represents the hosing flux, $F_R$ runoff (e.g. river outflow), $F_{P-E}$ the precipitation-minus-evaporation (e.g. rain, snow and evaporation), and $F_{\text{rest}}$ a minor rest term to close the budget consisting of calving, brine rejection and other processes. Moreover, for the precipitation-minus-evaporation it holds that
\begin{equation*}
    F_{P-E} = F_{\text{rain}} + F_{\text{snow}} +F_E
\end{equation*}
with rainfall $F_{\text{rain}}$, snowfall $F_{\text{snow}}$ and evaporation $F_E$. Note that we have omitted the hosing $F_H$ here. For the last term we have that
\begin{equation*}
    F_E = -\left(f_{\text{sic}}F_{E,\text{sic}} + (1-f_{\text{sic}}F_{E,\text{ocn}})\right)
\end{equation*}
where $f_{\text{sic}}$ is the fraction of surface that is covered by sea-ice, and $F_{E,\text{sic}}$ indicates evaporation via sublimation of sea-ice while $F_{E,\text{ocn}}$ indicates direct evaporation from the sea surface. As the former process is not as effective as the latter, evaporation is limited by sea-ice cover.

\subsubsection*{Software and model output}
The (processed) model output and analysis script is provided at \url{https://doi.org/10.5281/zenodo.16949317} and \url{https://doi.org/10.5281/zenodo.16949298}. CLIMBER-X is an open-source fast Earth system model, and can be found at \url{https://github.com/cxesmc/climber-x}.


\subsection*{Freshwater forcing for backward hysteresis simulations}
\begin{figure}
    \centering
    \includegraphics[width = \textwidth]{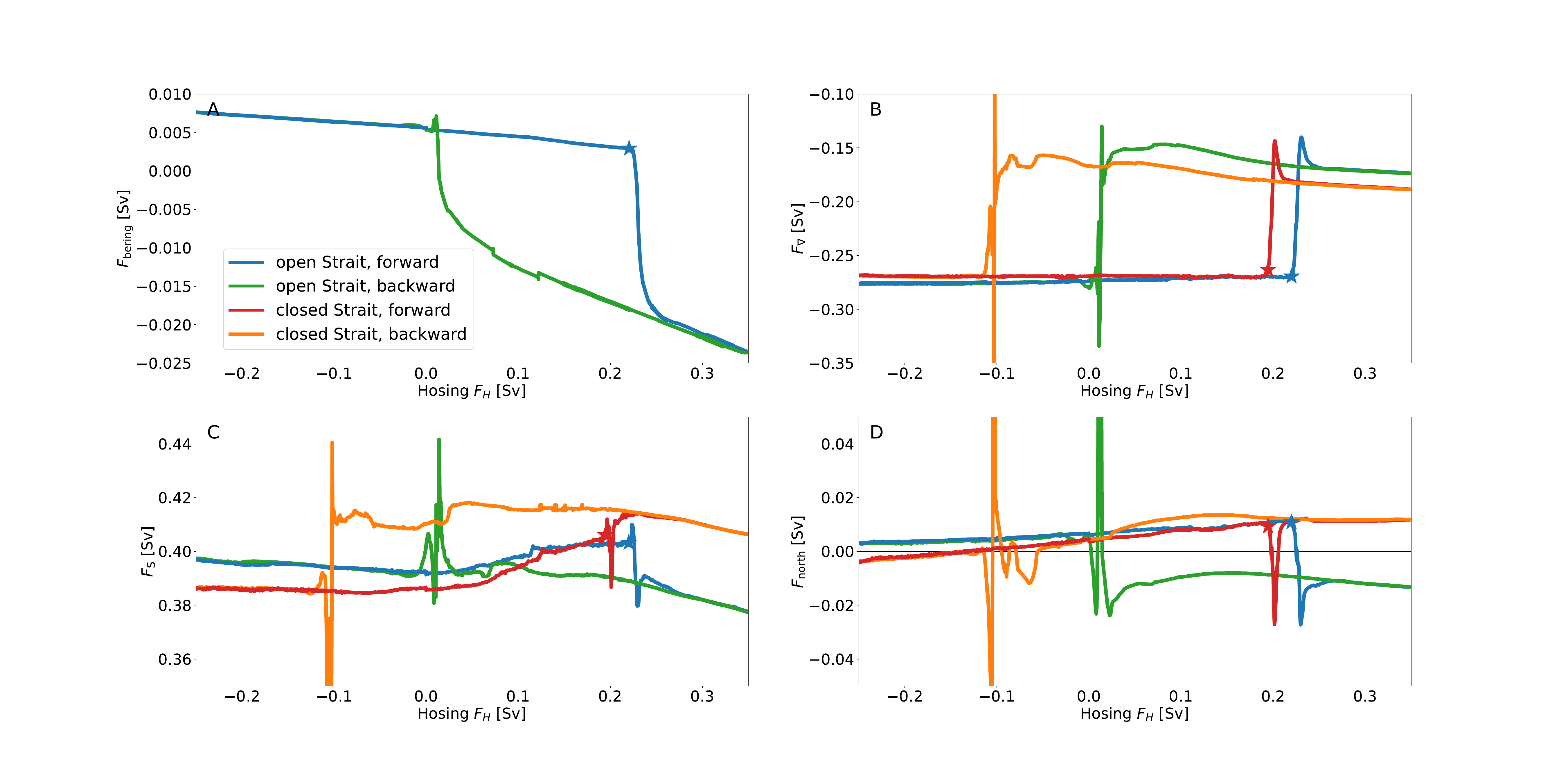}
    \caption{\textbf{The freshwater transports for complete hysteresis experiment.} Freshwater transports for the quasi-equilibrium simulations for an open Strait (blue, green), and a closed Strait (red, orange), consisting of simulations where the hosing flux $F_H$ increases (blue, red) and decreases (green, orange) (\textbf{A}-\textbf{D}). The asterisks mark the estimated tipping points of the AMOC collapses (\textbf{A}-\textbf{D}). (\textbf{A-D}) The freshwater transports through, respectively, the Bering Strait ($F_{\text{bering}}$), the lateral boundaries of the North Atlantic region ($F_{\nabla}$), the surface of the North Atlantic region ($F_S$), and the northern boundary of the North Atlantic region ($F_{\text{north}}$).} 
    \label{fig:supfwtransport}
\end{figure}

Regarding the quasi-equilibrium simulations for decreasing hosing (i.e. with mainly a collapsed AMOC), we see a more severe difference in freshwater exchanges than for the forward simulations, see also Figure~\ref{fig:supfwtransport}. For a collapsed AMOC with a closure there is a much larger freshwater transport southward out off the Arctic, resulting in a much fresher North Atlantic, see  Figure~\ref{fig:supSSS}\textbf{C}. Note however, that the North Pacific on the other hand is much more saline. With these larges discrepancies in sea surface salinities we also observe large differences in precipitation, evaporation and sea ice area between the OBS and CBS settings. In surface forcing to the North Atlantic the open Strait scenario exceeds that of a closed Strait, while the opposite is true for sea-ice export to it, see Figure~\ref{fig:supPmE}\textbf{B\&D}. Hence the much fresher sea surface under CBS must be caused by the larger southward freshwater and sea-ice transport from the Arctic. 

\newpage

\subsection*{Supplementary figures}
\begin{figure}[h!]
    \centering
    \includegraphics[width = \textwidth]{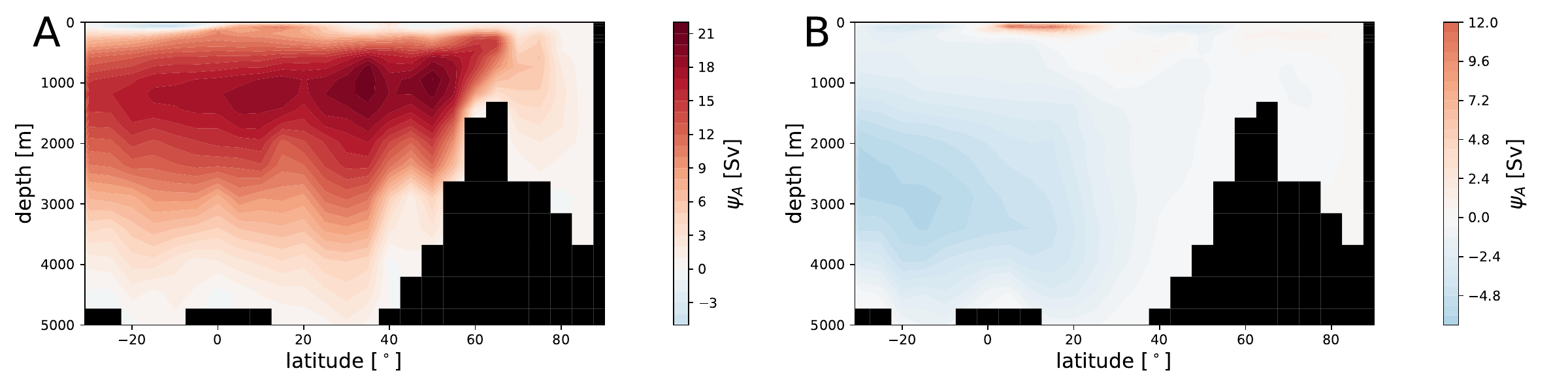}
    \caption{\textbf{The Atlantic overturning circulation.} The two equilibria of the AMOC at hosing $F_H = 0.1$~Sv and fixed CO$_2$ at 280~ppm, with the overturning streamfunction $\psi_A$ depicted in the Atlantic basin of an ON-state (\textbf{A}) and an OFF-state (\textbf{B}).}
    \label{fig:supAMOC}
\end{figure}

\begin{figure}
    \centering
    \includegraphics[width = \textwidth]{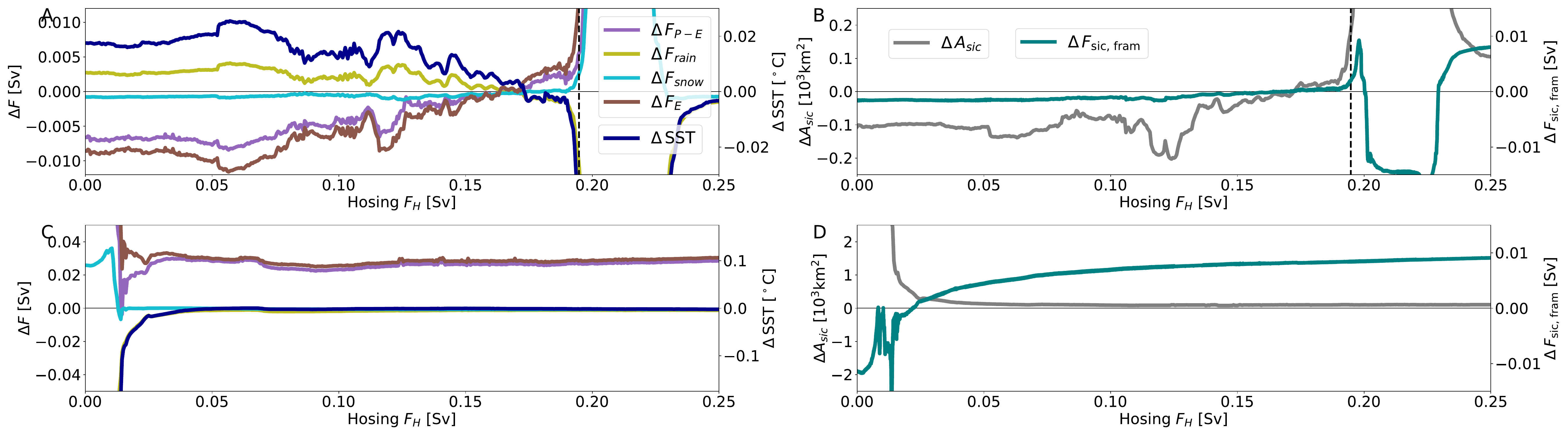}
    \caption{\textbf{Precipitation-minus-evaporation and sea-ice to the North Atlantic.} The difference between CBS and OBS conditions for the precipitation-minus-evaporation ($\Delta\,F_{P-E}$, purple), and its components rain ($\Delta\,F_{\text{rain}}$, citrus), snow ($\Delta\,F_{\text{snow}}$, purple), and evaporation ($\Delta\,F_{E}$, cyan) on the left axis, and for the average sea surface temperature (SST, blue) of the North Atlantic on the right axis, for the quasi-equilibrium simulations with (\textbf{A}) an active AMOC, and (\textbf{C}) a collapsed AMOC.  The difference between CBS and OBS conditions for the sea ice are in the North Atlantic ($\Delta\, A_{\text{sic}}$, gray) on the left axis, and for the southward sea ice export through the Fram Strait ($\Delta\, F_{\text{sic, fram}}$, teal) on the right axis, for the quasi-equilibrium simulations with (\textbf{B}) an active AMOC, and (\textbf{D}) a collapsed AMOC. The vertical dashed lines indicate the AMOC tipping point under CBS.}
    \label{fig:supPmE}
\end{figure}

\begin{figure}
    \centering
    \includegraphics[width = \textwidth]{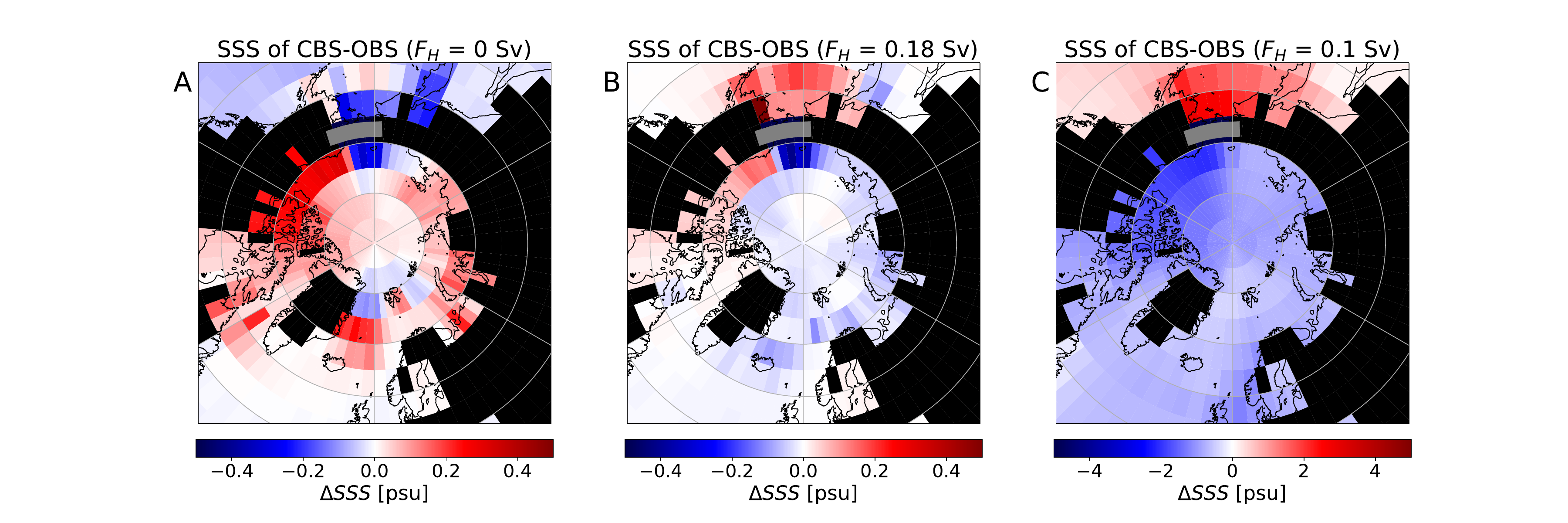}
    \caption{\textbf{Difference in sea surface salinities in the Arctic and North Atlantic.} The difference in sea surface salinities (SSS) in the Arctic and North Atlantic north of $55^\circ$N between CBS and OBS for the equilibria AMOC ON states at $F_H = 0$~Sv (\textbf{A}) and at $F_H = 0.18$~Sv (\textbf{B}), and for the equilibria OFF states at $F_H = 0.1$~Sv (\textbf{C}). The gray cells represent the BSD.}
    \label{fig:supSSS}
\end{figure}

\begin{figure}
    \centering
    \includegraphics[width = \textwidth]{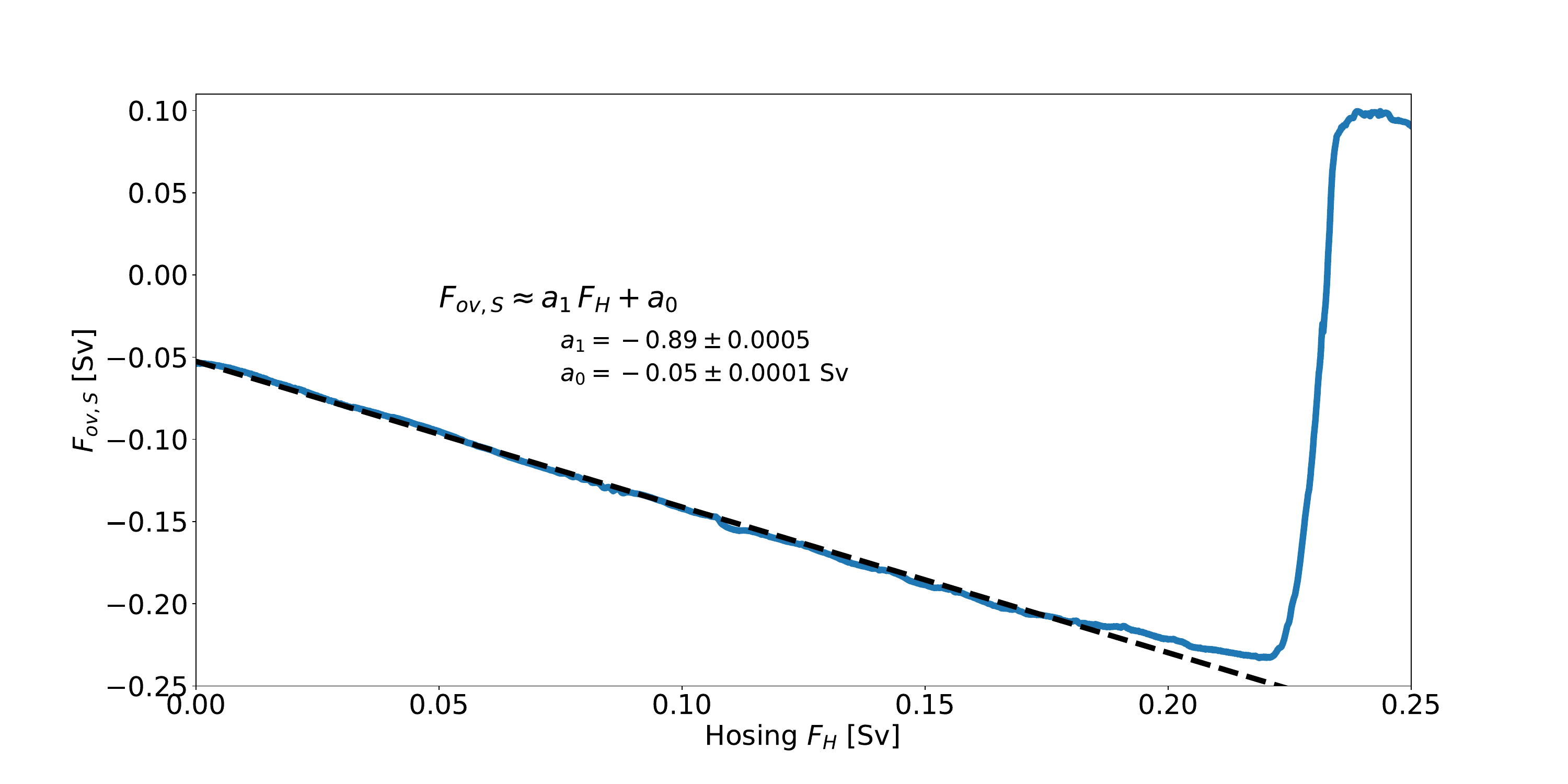}
    \caption{\textbf{Linear fit of $F_{ov,S}$ to the hosing.} The $F_{ov,S}$ (blue) of the quasi-equilibrium simulation starting in the AMOC ON equilibrium for OBS with hosing flux $F_H$ increasing at a rate $0.025$~Sv/kyr, together with a linear least-squares fit (black, dashed) to the data for range $F_H \in [0.00, 0.20]$~Sv.}
    \label{fig:supfit}
\end{figure}

\begin{figure}
    \centering
    \includegraphics[width = \textwidth]{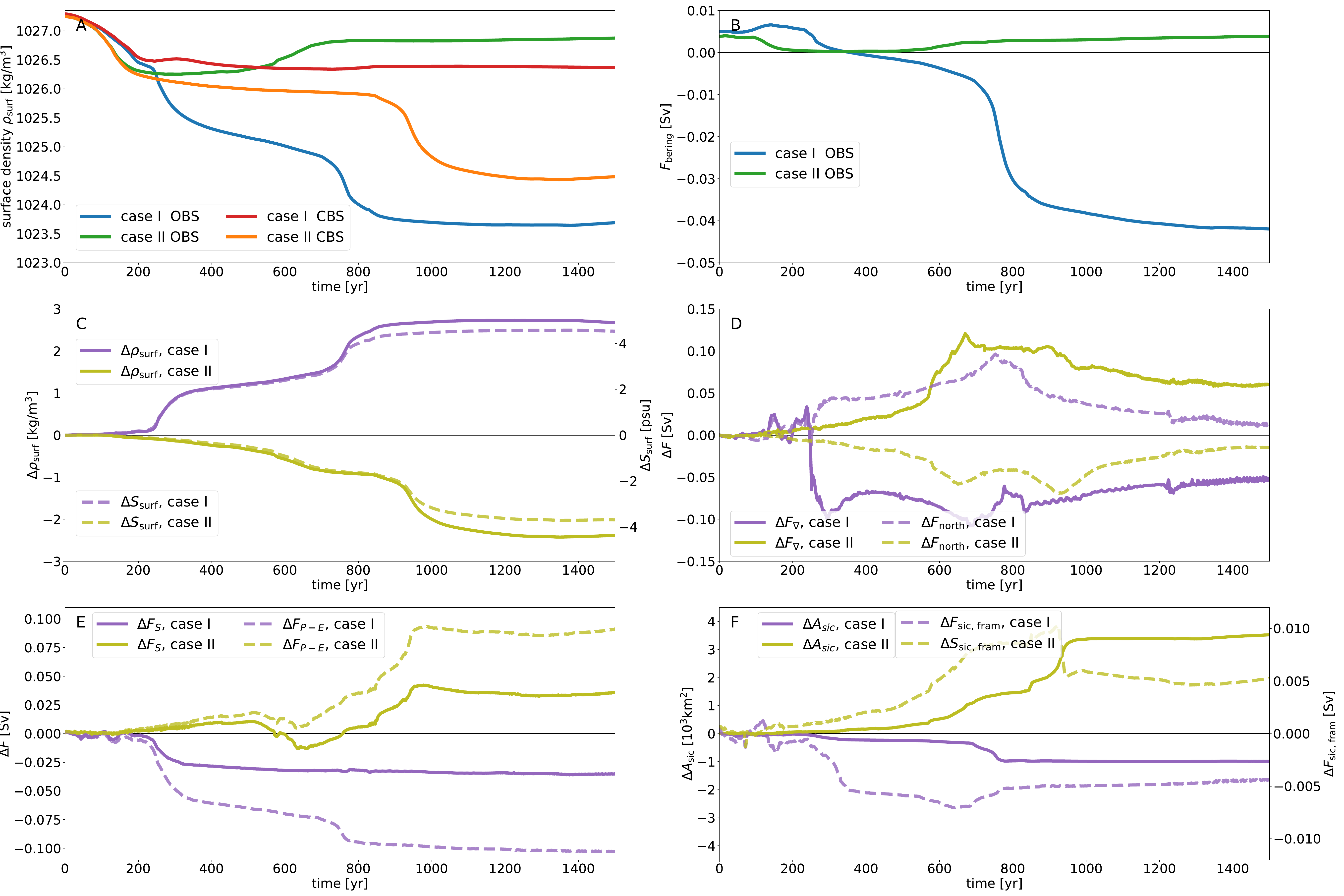}
    \caption{\textbf{Case I \& II diagnostics for complete simulation runtime.} Case I with a $1$~$\%$/yr CO$_2$ increase for $188$~yr and hosing $F_H = 0.05$~Sv with an open Strait (blue) and an immediate closure (red), and Case II with a $1$~$\%$/yr CO$_2$ increase for $93$~yr and hosing $F_H = 0.15$~Sv with an open Strait (green) and an immediate closure (orange) with their average density $\rho_{\text{surf}}$ of the top $200$~m of the North Atlantic region (\textbf{A}), and the freshwater transport through the Bering Strait $F_{\text{bering}}$ (\textbf{B}). Moreover, the difference between CBS and OBS settings for case I (purple) and case II (citrus) in surface density $\Delta\rho_{\text{surf}}$ (\textbf{C}, solid) and in surface salinity $\Delta S_{\text{surf}}$ (\textbf{C}, dashed), in freshwater import through the lateral boundaries $\Delta F_{\nabla}$ (\textbf{D}, solid) and in freshwater import through then northern boundary $\Delta F_{\text{north}}$ (\textbf{D}, dashed), in surface freshwater transport $\Delta F_S$ (\textbf{E}, solid) and in precipitation-minus-evaporation $\Delta F_{P-E}$ (\textbf{E}, dashed), and in sea-ice area in the North Atlantic $\Delta A_{\text{sic}}$ (\textbf{F}, solid) and in southward sea-ice export through the Fram Strait $\Delta F_{\text{sic, fram}}$ (\textbf{F}, dashed).}
    \label{fig:supcasediag}
\end{figure}

\begin{figure}
    \centering
    \includegraphics[width = \textwidth]{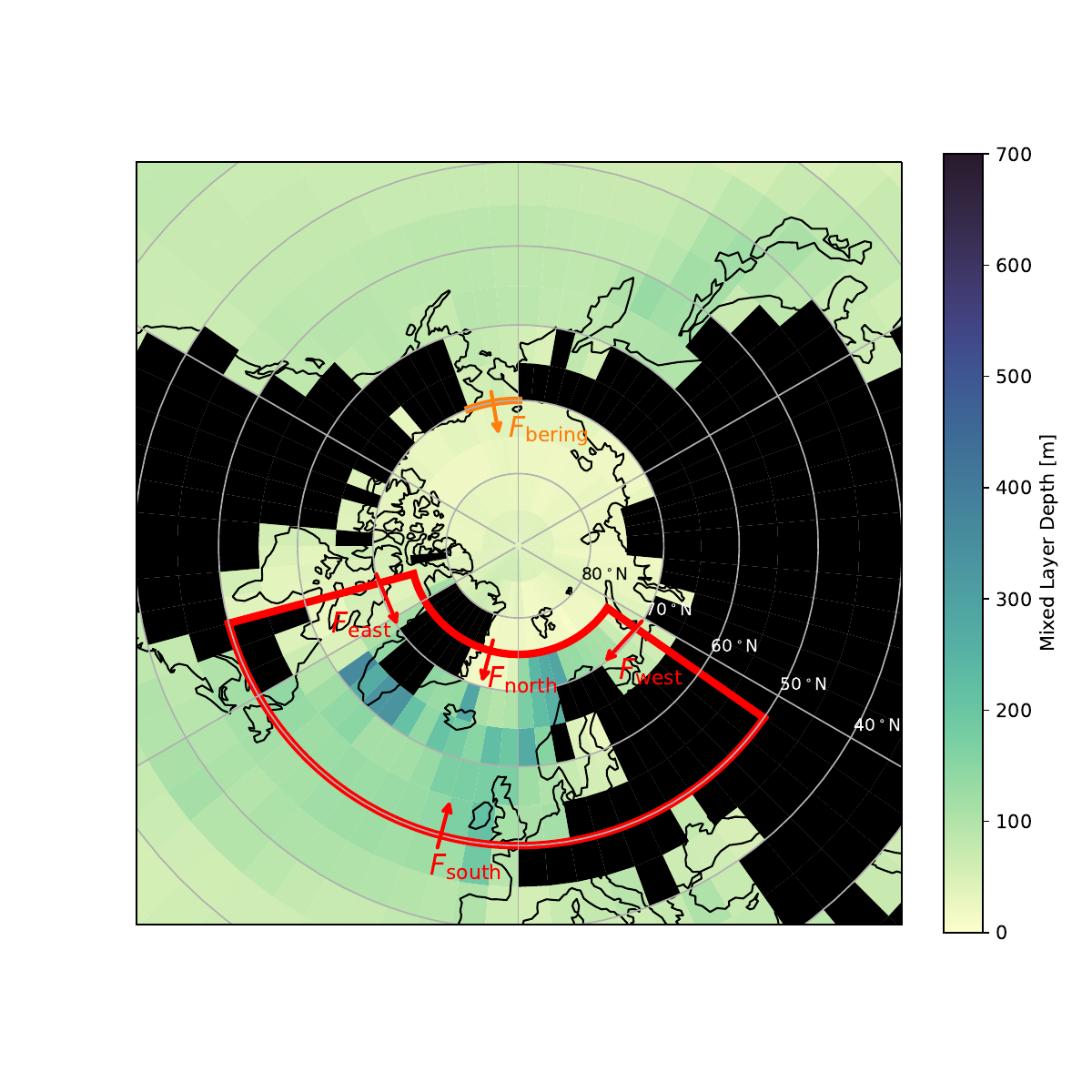}
    \caption{\textbf{The North Atlantic.} The selected North Atlantic region (enclosed in red) between latitudes $50^\circ$N and $75^\circ$N, and longitudes $75^\circ$W and $55^\circ$E, with the arrows indicating the direction of the computed freshwater transports through each boundary. The orange line indicates the section through which $F_{\text{bering}}$ is computed, with the arrow indicating its direction. The yearly-average mixed layer depth of an active AMOC for pre-industrial settings is shown. The black cells indicate grid cells with a zero ocean fraction, on top of the current coastlines (black, solid).}
    \label{fig:supregion}
\end{figure}





\end{document}